\documentstyle[12pt]{article}
\begin{document}
\newcommand{\be}{\begin{equation}}
\newcommand{\ee}{\end{equation}}
\large
\vskip 2cm
\begin{center}
{\Large {\bf Spin and New Physical Field \\}} \vspace*{1cm}
{\bf Ivanhoe B. Pestov \\} \vspace*{2cm}
{\bf Joint Institute for Nuclear Research
 \\ Bogoliubov Laboratory of Theoretical Physics\\
141980, Dubna, Moscow Region, Russia} \\ {E--mail: pestov@thsun1.jinr.ru}
\end{center}

PACS: 04; 11
\begin{abstract}
It is shown that the interactions between the fermion and the
gravitational fields are due to the torsion field. The torsion
field  is  considered to be  a potential one, like  the electromagnetic and
gravitational fields.  The field equations are obtained, which describe
the interactions of the torsion field with the conventional physical
fields.  The general covariant Lagrangian of the gravitational field, based
on the torsion field, is derived.  Experiment is proposed to test the
theory.   \end{abstract}


\section*{Introduction}
It is commonly assumed that the Einstein metric potential of the
gravitational field is a part of all equations of theoretical physics and
thus the gravitational field affects all the physical processes. This
statement (currently known as the principle of  universality of
gravitational interactions) has a fundamental significance for
understanding the structure of the physical notions and
laws, which forms a framework for all the physical processes.
The relativistic wave equation for the  electron
was suggested by Dirac in 1928. The question naturally arises how to
reconcile his equation with General Relativity.  This problem was
investigated in the works of Fock and Ivanenko [1] and Weyl [2].  The main
result of this consideration is that Einstein's gravitational
potential does not enter into the Dirac equation.  Thus, in the
microphysical realm the principle of universality of gravitational
interactions met with the difficulties of principal character that are not
overcome till now for the following reasons.   To save the
principle of universality of gravitational interactions it was suggested
that a true gravitational potential is a field (vierbein or tetrad) that
defines the general covariant form of the Dirac equation.  However, it is
quite evident that the validity of this assumption is not indisputable from
the logical point of view.  Indeed, we have no reasons to doubt that
Einstein's gravitational potential is a primary notion and hence  tetrads
may be (and should be ) incorporated into the framework of Einstein's
gravity theory  among the other  possible origins of the gravitational
field with proper energy--momentum tensor.
It is quite clear, but so far as we know  such formulation of the problem was
not considered in literature.  The main reason for this is that the
principle of universality of gravitational interactions is related
to another very interesting problem, namely the existence of a new
physical field due to spin.

With the discovery of the spin (spin of the electron) it was suggested
that  spin is the source of a new field (known in our
days as a torsion field).  The notion of torsion is causally related to
asymmetric connection which was mentioned by Eddington who pointed out on
some applications of this idea in microphysics. Torsion as antisymmetric part of
asymmetric affine connection was introduced by Cartan. He hinted
that the torsion might be connected with the intrinsic angular momentum of
matter.  Later the different
aspects of the new concept were developed in different directions in the
enormous number of publications ( see, e.g., the  original papers
of Eddington and Cartan [3], [4] and [5],[6] for extensive review
and references on various aspects of the theory with torsion field). But
despite all this efforts the physical meaning of the new field and its
possible link with spin is still an open problem.

Thus, spin and gravity are related to the  two actual
problems.  One of them deals with the description of the spinor fields
in framework of the metrical theory of gravity  and the other  with the
existence of a new physical field tightly connected with the
quantum--mechanical notion of spin.  In this paper we show that both
problems are interrelated and on this ground we propose their simple
solution.  The essence of our approach  is based on the assumption that a
spinning matter interacts with the gravitational field indirectly and the
torsion field is an interface between the gravitational and fermion fields.
Of course, this means that the principle of universality of gravitational
interactions does not hold valid anymore and that is why an experiment is
proposed to decide between the metrical and tetrad theories of gravity.

\section{The Dirac equation in general covariant form}
  Let ${\bf C}^4 $ be a linear space of columns of four complex
numbers $\psi_{1}, \,\psi_{2},\, \psi_{3},
\,\psi_{4}.$ Linear transformations in this space can be presented by
the  complex matrices
$(4 \times 4).$ The set of all invertible $(4 \times 4)$ complex matrices
forms a group denoted by $GL(4,{\bf C}).$ Dirac's  $\gamma^{\mu}$ matrices
belong to $GL(4,{\bf C}) $ and obey anticommutation relations  $$
\gamma^{\mu}\gamma^{\nu}+ \gamma^{\nu}\gamma^{\mu} = 2 \eta^{\mu \nu},$$
where  $\eta^{\mu \nu}$ are structure constants of the Poincar\' e
group, $\eta^{\mu \nu} = \eta_{\mu \nu} = {\mbox diag}(1,-1,-1,-1).$ From
$\gamma^{\mu}$  one can construct sixteen linear independent matrices that
form a basis of the Lie algebra of $GL(4,{\bf C}).$ This basis is
especially important since the matrices $S_{\mu \nu} = \frac{1}{2}
\gamma^{\mu}\gamma^{\nu} $ form the basis of the Lie algebra of the Lorentz
group (subgroup of  $GL(4,{\bf C}).$) Thus, a spinor is an element of the
space ${\bf C}^4.$  For better understanding it should be noted that in the
space ${\bf C}^3 $ there are no matrices like $\gamma^{\mu}.$

If one considers $\psi_{1}, \,\psi_{2},\, \psi_{3}, \,\psi_{4}$
as a set of complex scalar fields on the space--time manifold then
a spinor field emerges on the manifold which is a basis of irreducible
representation of the group $GL(4,{\bf C}).$  It is not difficult to
understand that $GL(4,{\bf C}) $  is a group of internal symmetry since
its transformations touch only functions of the spinor field and do not
affect the  coordinates. In other words, spin symmetry is {\bf an internal
symmetry}.

Now we consider general covariant formulation of the Dirac
equation in the Minkowski space--time. We shall follow the fundamental
physical principle that physical laws, like geometrical relationships,
depend neither on the choice of the coordinate system nor on the choice
of the basis in the studied vector spaces. With respect to an arbitrary
curviliner system of coordinates Minkowski space--time is characterized by
the metric $$ ds^2 = g_{ij} dx^i dx^j$$ of the Lorentz signature, which
satisfies the equation $R_{ijkl} =0.$ At given $g_{ij},$ the generators of
the group of space-time symmetry can be presented as a set of linear
independent solutions of general covariant system of equations (Killing's
equations)$$ P^i \partial_i g_{jk} + g_{ik} \partial_j P^i + g_{ji}
\partial_k P^i =0 $$ for a vector field $P^i.$  In the case of the
Minkowski metric we have ten the linear independent solutions of the
Killing equations, which are denoted $P^i_{\mu}$  and $M^i_{\mu \nu} = -
M^i_{\nu \mu} $ and  hence the Greek indices  enumerate a vector
fields and take the values $0,1,2,3,$ like  coordinate Latin indices.

It is well--known that the generators of the Poincar\'e group
$$ {\bf P_{\mu}} = P^i_{\mu} \frac{\partial}{\partial x^i}, \quad
  {\bf M}_{\mu \nu} = M^i_{\mu \nu} \frac{\partial}{\partial x^i} $$
satisfy the following commutation relations
\begin{equation} [{\bf{P}}_{\mu}, {\bf{P}}_{\nu}] =0, \label{1}
\end{equation}
\begin{equation} [{\bf{P}}_{\mu}, {\bf{M}}_{{\nu}{\lambda}}]
 = \eta_{\mu \nu} {\bf{P}}_{\lambda} - \eta_{\mu \lambda} {\bf{P}}_{\nu}.
 \label{2} \end{equation}
It is evident that  all these relations are
general covariant and that the operators
  ${\bf P}_{\mu}=  P^i_{\mu} \frac{\partial}{\partial x^i} $
transform a scalar field into the scalar one.

Now we shall show that the general covariant Dirac equation has the form
 \begin{equation}
		    i\gamma^{\mu} {\bf{P}}_{\mu} \psi = \frac{mc}{\hbar }
 \psi, \label{3}   \end{equation}
 where  $\psi$ is a column of four complex scalar fields in
question and ${\bf {P}}_{\mu} $ are the generators of
space--time translations.  Dirac's equation is covariant with respect to
the general coordinate transformations.  It is also clear that the equation
(3) is equivalent to the equation $$ i{\tilde {\gamma}}^{\mu} {\bf
{P}}_{\mu} \psi = \frac{mc}{\hbar } \psi,$$  if $ {\tilde {\gamma}}^{\mu} =
S \gamma^{\mu} S^{-1},$ where $S \in  GL(4,{\bf C}) $ (the equation (3) is
covariant with respect to the transformations of the group $ GL(4,{\bf C})
$ ). Since Dirac's equation (3) is general covariant we can use any system
of coordinates. In the preferred system of coordinates the equation (3) has
a customary form.

Now we have found enough to provide some valuable insights into the
connection between the space--time and internal transformations.
Consider again the generators of the internal Lorentz group
$S_{\mu \nu} = \frac{1}{4}( \gamma_{\mu} \gamma_{\nu} -
\gamma_{\nu} \gamma_{\mu}) $ and pay attention to the
commutation relations
\begin{equation} [\gamma_{\mu}, S_{{\nu}{\lambda}}] = \eta_{\mu \nu}
\gamma_{\lambda} - \eta_{\mu \lambda} \gamma_{\nu} \label{4}.
\end{equation} Comparing (2) and (4) it is not difficult to verify
that the operators  $${\bf{L}}_{\mu
\nu} = {\bf{M}}_{\mu \nu} + S_{\mu\nu} $$ commute with the Dirac
operator $D = i\gamma^{\mu} {\bf{P}}_{\mu} $  and satisfy  the commutation
relations of the Poincar\'e group. Thus, in the Minkowski space--time
there is a relation between the internal symmetry group and the space--time
symmetry group. The consequence is that Dirac's equation (3) is invariant
with respect to the transformations of the Poincar\'e group.

\section{Generalization of the Dirac theory }

Our goal here is to consider natural extension of the Dirac theory assuming
that now $P^i_{\mu}$ are simply components of the four linear
independent vector fields. In this case there is no strict relation
between the internal and space--time symmetry  but instead a
new possibility is opened.  We keep the internal symmetry on the
same ground as in the starting situation but attempt to
expand  the space--time symmetry to the diffeomorphism group.  It
may be possible if we can establish a system of equations for the fields
$\psi$ and $P^i_{\mu}$  that is invariant with respect to the
transformations of the diffeomorphism group.

Idea of physical interpretation of the new field  $P^i_{\mu}$ can
be derived from the principle of universality of  gravitational
interactions or from the other arguments which will be considered in the
following section.  If we claim that the principle of
universality of gravitational interactions is held valid, then we should
demand that a quadruplet of the linear independent vector fields is a
potential of the gravitational field.  Furthermore, we suggest that a new
gravitational potential enters into the Dirac equation in the form
considered above and into the equations of other fields in the form of
the metric tensor associated with a new gravitational potential $$ {\tilde
g}_{ij} = \eta_{\mu\nu} P^{\mu}_i P^{\nu}_j , $$ where $\eta_{\mu\nu}$
are the structure constants of the Poincar\'e group and $P^{\mu}_i $
is also a quadruple of covector fields, which is inverse to  $P^i_{\mu} $
\begin{equation} P^{i}_{\mu} P^{\mu}_j = \delta^i_j, \quad P^{i}_{\mu}
P^{\nu}_i =\delta^{\mu}_{\nu} \label{5}.  \end{equation} Taking into
account this postulate (which is alternative to the original Einstein's
idea) we need to derive the field equations for the new
gravitational potential.  There are two distinct approaches to this
problem.  One of them follows the line that equations for the $P^{i}_{\mu}
$  are defined completely by an associative metric $\tilde g_{ij}$, i.e.
in the Einstein-Hilbert action we simply replace  $g_{ij}$ by  $\tilde
g_{ij}$.  However, this postulate indicates that a theory
of the new field $P^{i}_{\mu} $   should be invariant with respect to the
local transformations of the form $$ {\bar P}^{i}_{\mu} = L^i_k P^{k}_{\mu}
, $$ where $L^i_k$  is a tensor field of the type (1,1), such that  $$L^i_k
L^j_l {\tilde g}_{ij}  = {\tilde g}_{kl}.$$  This condition is fulfilled
in some modification of the Dirac equation (see [1],[2], [7],[8]). The
second approach is based on the other Lagrangian and characterized by
the absence of local symmetry group defined above. In this case Dirac's
equation (3) needs no  modification ( See for example [9],[10] and further
references therein).

For the clarity we should emphasize that there are three successful
alternative theories of gravity. The first theory, originated by Einstein,
has a metric tensor as the basic structure. The problem (which is not solved
till now) is how to incorporate Dirac's theory into the framework of this
description of the gravity physics.   In the second and third approaches
the last problem is solved by  the method, which is not faultless from a
logical point of view.  Our goal here is to formulate a theory of
interactions of the gravitational field with the spinor field in the
framework of the Einstein metrical theory of gravity.

Thus, in the present work we state that a quadruplet of linear
independent vector fields is a new physical field,
which  should be  incorporated into the framework of the Einstein
gravity theory as an origin of the gravitational field with a proper
energy--momentum tensor.  Below we shall consider this possibility in more
details.  Of course, this idea leads to the violation of the
principle of universality of gravitational interactions but not to a
modification of the theory of the gravitational field ( known as
General Relativity ).

Now it is time  to consider a geometrical meaning of the new field
$P^i_{\mu}.$  As it is well known the Einstein gravitational potential
has a simple geometrical interpretation as an element of length in the
Riemann space--time.  In 1917 Levi--Civita  proposed how to introduce a
parallel transport as an internal notion on the Riemann manifold and opened
a way for generalizations.

Consider a vector field $V^i(x).$  Equation of local parallel transport
from a point $x^i$ to a point $ x^i+dx^i $  has in general the form
 \be  dV^i(x) = - L^i_{jk}(x) V^k(x)
dx^j \label{6} ,\ee where  functions $L^i_{jk}(x) $ are
components of a new geometrical object on the manifold, called a linear or
affine connection  $L$.  Under a parallel transport along the
infinitesimal close curve the change  of the vector is equal to the quantity

$$	   \triangle V^k = - B_{ijl}^k V^l dx^i \delta x^j,  $$
where
\be
   B_{ijl}^k = \partial_i L^k_{jl} -  \partial_j L^k_{il} +
L^k_{im} L^m_{jl} - L^k_{jm} L^m_{il} \label{7}
\ee
is a tensor field of the type (1,3), called the Riemann tensor of
the affine connection $ L^i_{jk} .$

From (6) it follows that under a coordinate mapping
 $$\bar x^i = \bar x^i(x), \quad x^i = x^i(\bar
x),$$ the transformation law for a  $ L^i_{jk} $    has the form
 \begin{equation} \bar L^i_{jk} = \frac{\partial
\bar x^i}{\partial x^l} ( L^l_{mn} \frac{\partial x^m}{\partial \bar x^j}
\frac{\partial x^n}{\partial \bar x^k} + \frac{\partial^2 x^l}{\partial
\bar x^j \partial \bar x^k } ) \label{8}.  \end{equation}
We shall say that a geometrical quantity is irreducible if it is possible
to find linear combinations of its components which themselves constitute a
new geometrical quantity. It is very important that under the coordinate
mappings a linear connection is a reducible quantity.  It is easy to see
from the expansion $$ L^i_{jk} =\frac{1}{2} (L^i_{jk} + L^i_{kj}) +
\frac{1}{2} (L^i_{jk} - L^i_{kj}).$$ From (8) it follows that a symmetrical
part of the affine connection $$ \{^i_{jk} \} = \frac{1}{2} (\Gamma^i_{jk}
+ \Gamma^i_{kj}) ,$$ is again the affine connection and the antisymmetrical
part, \begin{equation} H^i_{jk} = \frac{1}{2} (\Gamma^i_{jk} -
\Gamma^i_{kj}) \label{9}, \end{equation} transforms  as a tensor
field of  the type (1,2). This tensor field is called the torsion tensor.
Thus, we have two independent irreducible fields. It is very important that
with respect to the first field (symmetric affine connection) a metric can
be considered as potential of this new field in accordance with the
relation \begin{equation} \{^i_{jk} \} = \frac{1}{2} g^{il}(\partial_j
g_{kl} + \partial_j g_{kl} -\partial_l g_{jk} ) \label{10}.\end{equation}
Expression on the right hand side (10) was first found by
Christoffel.  For comparison,  the formula $$F_{ij} = \partial_i A_j -
\partial_j A_i $$ provides the link between a vector potential of the
electromagnetic field and a tensor of this field.  If we take a tensor
field (for example $g_{ij}$) and form its derivatives ($\partial_i g_{jk}$
) then these derivatives are neither the components of a tensor of any
geometrical object. But if one can form new geometrical object from these
derivatives then this process can be called natural derivative.  For
example, symmetrical affine connection can be considered as a natural
derivative with respect to the metric tensor and a bivector of the
electromagnetic field  as a natural derivative of the covector field. With
this transparent presentation we can put forward the
idea that torsion field (being new physical field) should be (like the
electromagnetic and the gravitational field) a potential field. Now the
clear problem is to show that the torsion tensor is a natural derivative
with respect to the unknown field (potential of the torsion field). We
state that a quadruplet of linear independent vector fields forms the
potential of the torsion field because \be H^i_{jk} =
P^i_{\mu} (\partial_j P^{\mu}_k - \partial_k P^{\mu}_j). \label{10} \ee So,
a torsion tensor is actually a natural derivative of the
$P^i_{\mu} .$

Since curvature and torsion are tightly connected then it is
natural to suppose that a geometrical  Lagrangian for these fields has
the form $L= B,$  where $B$ is a scalar  that can be constructed
from the Riemann tensor (7).   Indeed,
from (7), (9), (10), (11) it follows that \be B^l_{ijk}= R^l_{ijk} +
\nabla_i H^l_{jk} - \nabla_j H^l_{ik} + H^l_{im} H^m_{jk} - H^l_{jm}
H^m_{ik} \label{11}, \ee where \be R^l_{ijk}= \partial_i \{^l_{jk}\} -
\partial_j \{^l_{ik}\} + \{ ^l_{im}\} \{^m_{jk}\} - \{^l_{jm}\} \{^m_{ik}\}
\label{12} \ee is the well--known curvature tensor and $\nabla_i$ stands
for the covariant derivative with respect to the connection (10),
$$\nabla_i H^l_{jk} = \partial_i H^l_{jk} + \{^l_{im} \} H^m_{jk} -
\{^m_{ij} \} H^l_{mk} - \{^m_{ik} \} H^l_{jm}.$$ By contraction we derive
from (12) a more simple tensor \be B_{jk} = B^i_{ijk}= R_{jk} + \nabla_i
H^i_{jk} - \nabla_j H^i_{ik} + H^i_{im} H^m_{jk} - H^i_{jm} H^m_{ik}
  \label{13}, \ee where  $R_{jk}$ is the Ricci tensor. From (14) it follows
that a required scalar  is  $$ B = g^{jk}B_{jk}= R + g^{jk} H^l_{jm}
H^m_{jl} - \nabla_j H^j, $$ where $R$ is scalar curvature (the
Einstein-Hilbert Lagrangian of the gravitational field), and $H^j = g^{jk}
H^i_{ik}.$ We see that the Lagrangian of the torsion field itself is given
by the formula \be L_t = \frac{1}{2} g^{jk} H^l_{jm} H^m_{kl} \label{15}.
\ee

As a byproduct of our consideration we shall derive now a new general
covariant form of the Lagrangian of the gravitational field. Motivation for
our study is well--known ( The Einstein-Hilbert Lagrangian $R$
contains a second order derivatives of $g_{ij}$ and this leads to the known
difficulties [11]).

Consider a binary tensor field
\be B^i_{jk} =  P^i_{\mu} \nabla_j P^{\mu}_k =  \Gamma^i_{jk}- \{
^i_{jk}\} \label{16},  \ee where
\be	 \Gamma^i_{jk}	=  P^i_{\mu} \partial_j P^{\mu}_k \label{16}  \ee
is the canonical connection for which the Riemann tensor is equal to zero
identically.
Setting $$\Gamma^i_{jk}=\{ ^i_{jk}\} +\Gamma^i_{jk}- \{ ^i_{jk}\}
=\{ ^i_{jk}\} +B^i_{jk} $$ and following  closely the line defined by
(12), (13), (14), we derive the relation $$ 0 = R_{jk} + \nabla_i B^i_{jk}
- \nabla_j B^i_{ik} + B^i_{im} B^m_{jk}- B^i_{jm} B^m_{ik}.$$ From the
last formula it follows that  $$ R + \nabla_i(g^{jk}
B^i_{jk} - g^{ik} B^l_{lk})= g^{jk} ( B^i_{jm} B^m_{ik} - B^i_{im}
B^m_{jk}).$$ Thus, the Einstein-Hilbert Lagrangian is equivalent to
the Lagrangian $$L_q =\frac{1}{2} g^{jk} ( B^i_{jm} B^m_{ik}
- B^i_{im} B^m_{jk}) .$$
This Lagrangian is based on the torsion field and may be more convenient in
the quantum theory of the gravitational field.

\section{Field equations for the torsion field}
We make small variations in our field quantities $P^i_{\mu} $ and
calculate the change in the action integral
$$ A =   \frac{1}{2} \int g^{jk} H^l_{jm} H^m_{kl} \sqrt g d^4x$$
with $g = - Det(g_{ij}).$
It is convenient to introduce tensor $$F^{ij}_k =g^{il} H^j_{lk} - g^{jl}
H^i_{lk} = H^{ij}_k - H^{ji}_k $$ with inverse transformation $$H^i_{jk} =
\frac{1}{2}(g^{il} F^{mn}_l g_{jm} g_{kn}  + g_{jl} F^{il}_k - g_{kl}
F^{il}_j ). $$
Since  $$ H^i_{jk} = P^i_{\mu} (\partial_j P^{\mu}_k - \partial_k
P^{\mu}_j )= P^i_{\mu} (\nabla_j P^{\mu}_k - \nabla_k P^{\mu}_j ) , $$
we get sequentially
\be \delta L_t = F^{jk}_l \delta(P^l_{\mu} \nabla_j
P^{\mu}_k) = F^{jk}_l (\nabla_j P^{\mu}_k) \delta P^l_{\mu} + F^{jk}_l
 P^l_{\mu} \nabla_j \delta P^{\mu}_l \label{18}  .  \ee With (5) $$
 \delta P^{\nu}_k = - P^{\nu}_l P^{\mu}_k \delta P^l_{\mu}. $$ By
 this, the second term in the right hand side of
 (18) can be presented in the following form \be \nabla_j
 (F^{jk}_l P^l_{\mu} \delta P^{\mu}_l) + P^{\mu}_k (\nabla_j F^{jk}_l +
 F^{jk}_m P^{\nu}_l \nabla_j P^m_{\nu} ) \delta P^l_{\mu} \label{19}.  \ee
 From (18) and (19) it follows that the variational principle provides the
 following equation for the potential of the torsion field
 \be
 P^{\mu}_k \nabla_j F^{jk}_l + F^{jk}_l \nabla_j P^{\mu}_k + F^{jk}_m
 P^{\mu}_k P^{\nu}_l \nabla_j P^m_{\nu} = 0 \label{20}.  \ee It
 is possible to rewrite this equation in various forms.
 With (5) and (16) we have $$ P^{\nu}_l \nabla_j P^m_{\nu}= -P^m_{\nu}
 \nabla_j P^{\nu}_l= - B^m_{jl}, \quad \nabla_j P^{\mu}_k= B^m_{jk}
 P^{\mu}_m   $$ and hence (20) can be written in the following form
  \be \nabla_j
 F^{jk}_l  + B^k_{jm} F^{jm}_l - B^m_{jl} F^{jk}_m = 0 \label{21}.  \ee
 Let
 $\stackrel{0} {\nabla}_i$ be a covariant derivative with respect to
 the canonical connection
 (17).  Since $$\nabla_j F^{jk}_l = \stackrel{0} {\nabla}_j F^{jk}_l
 -B^j_{ji} F^{ik}_l -B^k_{jm} F^{jm}_l + B^m_{jl} F^{jk}_m,$$
 equations (21) can be presented as follows  \be
 (\stackrel{0} {\nabla}_j-B_j) F^{jk}_l =0 \label{22}, \ee where $B_i$
 is a trace of the binary tensor field (16), $B_i = B^k_{ki}.$

\section{Interactions with known physical fields}
In this chapter we shall consider interactions of the torsion field with
known fundamental fields, the fermion, the gravitational and
the electromagnetic.

Dirac's Lagrangian has the form
  \begin{equation} L_D =
 \frac{i}{2}P^i_{\mu} \biggl(\bar \psi \gamma^{\mu} D_i \psi - (D_i \bar
	   \psi)\gamma^{\mu} \psi\biggr)- m \bar \psi \psi \label{23} ,
\end{equation} where  $P^i_{\mu}$ is a potential of the torsion
field and $$ D_i \psi = (\partial_i - ie A_i) \psi,
\quad D_i \bar \psi = (\partial_i + ie A_i) \bar \psi $$ as
usually. Dirac's Lagrangian is invariant with respect to the substitutions
$$\psi \Rightarrow e^{i\lambda} \psi, \quad \bar \psi \Rightarrow
e^{-i\lambda} \bar \psi, \quad A_i \Rightarrow A_i + \partial_i \lambda. $$
Action has the form $$A = \int L_{D} \, p \,d^4x,$$ where $p = \mbox{Det}
(P^{\mu}_i).$ Since $$P^{i}_{\mu} \partial_j P^{\mu}_i = \frac{1}{p}
\partial_j p,$$  this action leads  to the Dirac equations in the
presence of  external torsion and electromagnetic fields \be i P^i_{\mu}
\gamma^{\mu} (D_i - \frac{1}{2} H_i) \psi = m \psi \label{24} ,\ee \be i
P^i_{\mu}  (D_i - \frac{1}{2} H_i){\bar \psi} \gamma^{\mu} = -m {\bar \psi}
\label{25} ,\ee where $H_i$ is a trace of the strength tensor of the
torsion field $H_i= H^k_{ki}.$

Setting $$ W^{\mu}_i = \frac{i}{2} (\bar \psi \gamma^{\mu}
D_i \psi - (D_i \bar \psi)\gamma^{\mu}  \psi),  $$ we have $L_D
=P^i_{\mu} W^{\mu}_i- m \bar \psi \psi .$  Hence, from the action
 $$A = \int L_{D} \, p \,d^4x +\int L_t \sqrt{g}
d^4x, \quad g= -\mbox{Det}(g_{ij}), $$ we derive (in accordance with
(21) ) the following equation for the torsion field
\be \nabla_j F^{jk}_l + B^k_{jm}
F^{jm}_l - B^m_{jl} F^{jk}_m + W^k_l =0 \label{26}, \ee where $$W^k_l =
\epsilon P^k_{\mu} W^{\mu}_l, \quad \epsilon = p/ \sqrt g.  $$ The
equations (24), (25) and (26) explain clearly how the torsion field
interacts with the spinor field.

From  the equation (26) an interesting relation can be derived. By
summing over the indices  k and l we get  that a trace of $H^i_{jk}$
satisfies the following equation  \be \nabla_i H^i  = m
{\bar \psi} \psi \label{27}, \ee where $H^i = g^{ik} H_k.$ We
conclude that for $m=0$ the interactions of the torsion and spinor fields
are characterized by a new conserved quantity. Indeed, this fact simply
means that the action is invariant under the mapping $$P^{\mu}_i
\rightarrow a P^{\mu}_i, \quad \psi \rightarrow a^{-\frac{1}{2}} \psi,$$
where $a $ is dimensionless constant.

Let
	$$L_M = - \frac{1}{4} F_{ij} F^{ij} $$
be a standard Lagrangian of the electromagnetic field. From the action
$$A = \int{L_M} \sqrt{g} d^4 x + \int{L_D} \,p
\,d^4 x $$ we derive  the equations of the electromagnetic field
 \be \nabla_i F^{ij}  + e J^i =0
, \quad J^i = \epsilon P^i_{\mu} {\bar \psi} \gamma^{\mu} \psi
\label{28}.\ee

We write the joint Lagrangian of the gravitational, torsion and
electromagnetic fields
\be L = \frac{1}{2} R + \frac{1}{2} H^k_{il} H^l_{jk} g^{ij} -\frac{1}{4}
F_{ij} F^{ij} = L_g + L_t + L_M \label{29}.\ee Varying action
 $$A = \int{L \sqrt g} d^{4}x  $$ with respect to  $g^{ij},$ we get
Einstein's equation \be G_{ij} = g_{ij} L_t -
H^k_{il} H^l_{jk} + F_{ik} F_{jl} g^{kl} + g_{ij} L_{M} \label{30} \ee
with the stress energy--momentum tensor of the torsion field
\be T_{ij} = g_{ij} L_t - H^k_{il} H^l_{jk} \label{31}. \ee  Notice that
the trace of energy--momentum tensor (31) is not equal to zero.

Thus, the basic equations of the torsion field interacting with the
known physical fields are derived. Now it is important to show that these
equations are compatible. To this end let us establish the identities for
the Lagrangians of the fields in question. For Dirac's Lagrangian one can
 derive the identity \be \partial_j L_D = D_j {\bar \psi} \frac{\delta
L_D}{\delta {\bar \psi}} - \frac{\delta L_D}{\delta {\psi}} D_j {\bar \psi}
+ \frac{1}{\epsilon}(\nabla_i W^i_j - B^i_{jk} W^k_i +e F_{ji} J^i)
\label{32}.\ee From (32) it follows that the circulation
of the energy of the spinning matter is defined by the equation
 \be \nabla_i W^i_j - B^i_{jk} W^k_i + eF_{ji} J^i = 0
\label{33},\ee when the electromagnetic and torsion fields are
present. The energy--momentum tensor $W^i_j$  of the spinning matter
 is not symmetric ( some interesting details of this phenomenon  and
further references  can be found in the review article [5]).

Identity for the Lagrangian of the torsion field may be written as follows
\be \partial_j L_t = \nabla_i S^i_j - B^i_{jk} S^k_i + \nabla^i
(H^k_{il}  H^l_{jk} )  \label{34},\ee
where
$$ S^i_j = \nabla_k F^{ki}_j  + B^i_{kl} F^{kl}_j - B^l_{kj} F^{ki}_l =
 P^i_{\mu} \frac{\delta L_t}{\delta {P^j_{\mu}}} . $$
It is necessary to illuminate  the important points under the
derivation of the identity (34). We have $$\partial_j L_t = F^{ik}_l
 (\nabla_j P^l_{\mu}) \nabla_i P^{\mu}_k + F^{ik}_l P^l_{\mu} \nabla_j
 \nabla_i P^{\mu}_k. $$ With Ricci's identity   $$\nabla_j \nabla_i
 P^{\mu}_k = \nabla_i \nabla_j P^{\mu}_k - R^l_{jik}P^{\mu}_l $$ we can
 represent the second term in the right hand side of first
relation in the following form $$ \nabla_i (F^{ik}_l P^l_{\mu} \nabla_j
P^{\mu}_k) -(\nabla_i (F^{ik}_l P^l_{\mu})) \nabla_j P^{\mu}_k - F^{ik}_l
R^l_{jik}.  $$ For the further transformations one needs to use identity
$$ F^{ik}_l R^l_{jik} = \nabla_i \nabla_k F^{ik}_j $$ and relations (5) and
(16).

From Einstein's equation (30) and identity (34)  it follows
that $$\nabla^i G_{ij} = \nabla_i S^i_j - B^i_{jk} S^k_i +
F_{jk} \nabla_i F^{ik}.$$ Since  $\nabla^i G_{ij} =0 $ identically, the
right hand side of the last equation should be equal to zero. From the
equations (26), (28) ¨ (33) it follows  that this is indeed the case.
Thus, the theory of dynamical torsion is described by the compatible
system of equations.
\section*{Conclusion}
Here we suggest an experiment to test the  formulated theory  and to
make choice between the alternative  theories of the gravitational field.
It is suggested to measure  the gravitational acceleration of electrons and
positrons in the Earth gravitational field. The motivation is as follows.

In 1967 Witteborn and Fairbank measured the net vertical component of
gravitational force on electrons in vacuum enclosed by a copper tube [12].
This force was shown to be less than 0.09 mg, where m is the inertial mass
of the electron and g is $980 cm/sec^2.$ They concluded
that this result supports the contention that gravity induces an electric
field outside a metal surface, of such magnitude and direction  that
the gravitational force on electrons is cancelled. If this is true, then
the positrons will fall in this tube with the acceleration $a = 2g.$ The
conclusion from the theory presented here is that electrons and positrons
do not interact with the gravitational field directly but only through the
torsion field.  And the result presented by the measurements may be
considered as an estimation for the energy of torsion field generated by
electron (and positron). Thus, the measurements  of the net vertical
component of the force on positrons in vacuum enclosed by a copper tube
will have the fundamental significance for understanding of the conceptual
basis of contemporary theoretical physics.

\newpage
\begin{flushleft}
{\bf References}
\end{flushleft}

 \end{document}